# Neutron-$^4$He Resonant Scattering at d-$^3$H Threshold

Bernard Hoop[a], Gerald M. Hale[b] and Petr Navratil[c]

[a]*Riley College of Education, Walden University, Minneapolis, MN 55401-2511*
[b]*Theoretical Division, Los Alamos National Laboratory, Los Alamos, NM 87545*
[c]*TRIUMF, 4004 Wesbrook Mall, Vancouver BC V6T 2A3, Canada*

**Abstract.** Speed plot and time delay analysis confirms a two-pole *S*-matrix structure of the lowest $3/2^+$ level in $^5$He. Real parts of neutron-$\alpha$ elastic and $^4$He(n,d)$^3$H reaction pole energies are 17.669+/−0.006 and 17.705+/−0.001 MeV n-$^4$He c.m., respectively. Neutron-$\alpha$ elastic resonance, uncorrected for experimental energy spread, is characterized by deuteron and neutron partial widths ca. 0.008 and 0.030 MeV suggested by previous time delay analysis.

## INTRODUCTION

Resonant elastic neutron-$^4$He scattering near the $^4$He(n,d)$^3$H reaction threshold and the large cross section of the inverse $^3$H(d,n)$^4$He reaction are two distinct features of the lowest $3/2^+$ (16.84-MeV) state in $^5$He at 17.64 MeV n-$\alpha$ c.m. [1]. This state is an n-$\alpha$ resonance on the second Riemann energy sheet of both n-$\alpha$ and d-$^3$H channels, with associated shadow pole on a different unphysical sheet that, through its associated zero on the physical sheet, contributes to the large $^4$He(n,d)$^3$H cross section. With reference to the $D_{3/2}$ resonant n-$\alpha$ scattering amplitude, multilevel *R*-matrix analysis suggests that n-$\alpha$ resonant phase shift increases through 90$^o$ [2], but single-level *R*-matrix analysis suggests that it does not [3]. *Ab initio* theory employing two-nucleon interaction and nine deuteron pseudo-states suggests that the resonant phase shift does not increase through 90$^o$ [4].

Multilevel *R*-matrix analysis [2] provides an *S*-matrix pole structure that characterizes this level as two poles: a main or conventional pole on the unphysical Riemann energy sheet nearest the physical sheet, where both n-$\alpha$ and d-$^3$H wave numbers have negative imaginary parts, and a shadow pole where the imaginary part of n-$\alpha$ wave number is negative, but d-$^3$H wave number is positive.

In this report, we employ speed plot and time delay analysis to elucidate the two-pole nature of this resonance.

## ASSUMPTIONS AND CALCULATIONS

D-wave neutron-$\alpha$ resonant elastic scattering cross section is approximated by a one-level Breit-Wigner model using deuteron and neutron partial widths taken from *R*-matrix analysis [2]. Non-resonant total n-$\alpha$ cross section in other partial waves is represented by a nuclear Ramsauer model with n-$\alpha$ interaction radius 3.3 fm derived from a best fit to n-$\alpha$ total cross section in the neutron energy range 8 to 16 MeV, where the 0$^o$ differential cross section is within 2% of Wick's limit [5]. Non-resonant $D_{3/2}$ phase shift is taken as scattering from a hard sphere potential of radius 2.013 fm. In the present study, cross section for the $^4$He(n,d)$^3$H reaction is calculated via detailed balance from the astrophysical S-factor for the inverse reaction and fitted by an empirical function [6]. Peak reaction cross section is held fixed, unlike in an earlier study, where it was allowed to vary [7].

Calculated elastic and reaction cross sections are added to yield total cross section and compared with all available n-$\alpha$ total and elastic scattering cross section data via an overall squared error of fit. Elastic cross sections derived from integrated angular distributions are multiplied by 1.1 to account for scale factor normalization uncertainty. Experimental neutron energy spread of 0.040 MeV is taken into account by convoluting a Gaussian energy distribution with calculated elastic cross section and re-evaluating with a Breit-Wigner model [7].

Real part of complex time delay Δt(E) and *Speed(E)* vs n-α c.m. energy E

$$\Delta t(E) = \hbar \, \text{Re}\left\{\frac{d}{dE}\left[\arg\left(\frac{S-1}{2i}\right)\right]\right\} = 2\hbar \frac{d\mu}{dE}$$

and

$$Speed(E) = 2\hbar \left|\frac{d}{dE}\left(\frac{S-1}{2i}\right)\right| = 2\hbar \sqrt{\left(\tau \frac{d\mu}{dE}\right)^2 + \frac{1}{4}\left(\frac{d\tau}{dE}\right)^2}$$

are determined from Argand diagrams of $D_{3/2}$ scattering amplitude derived from calculated total and elastic cross sections, where resonant scattering amplitude is $(S-1)/2i$, $S = \exp(2i\delta)$, complex resonant phase shift $\delta = \mu - i/2 \ln \tau$, $\mu = \text{Re}(\delta)$, and where $D_{3/2}$ inelastic parameter $\tau = \exp[-2\,\text{Im}(\delta)]$ is derived from calculated $^4$He(n,d)$^3$H cross section obtained as described above.

As outlined elsewhere [7], time delay is calculated from energy derivatives and compared with best fits to the sum of two Lorentzians representing conventional and shadow pole time delays plus n-α transit time. *Speed(E)* is similarly calculated and compared with the sum of two Lorentzians, except without n-α transit time term.

## RESULTS

Multilevel *R*-matrix analysis [2], where resonant phase shift increases through 90°, yields maximum elastic time delay and speed at 17.67 MeV. Maximum reaction time delay and speed occur at 17.70 MeV. Single level *R*-matrix analysis [3], where resonant phase shift does not increase through 90°, yield maximum elastic time delay and speed at less than 17.66 MeV, and maximum reaction time delay and speed, respectively, at 17.69 and 17.71 MeV. *Ab initio* theory using NN interaction and nine deuteron pseudo-states [4] yields maximum elastic time delay and speed at 17.68 MeV. Maximum reaction time delay and speed occur at 17.70 MeV.

An improvement upon the above determinations of the two pole energies is achieved by calculating time delay and speed within a Breit-Wigner model for n-α elastic resonance peak energy varied systematically in 0.002-MeV steps from neutron energy 22.110 to 22.190 MeV (17.659 to 17.723 MeV n-α c.m.). Partial width (0.5$\hbar$ divided by maximum speed) was found to be largest at 17.674 and 17.705 MeV. Maximum elastic time delay and speed for three representative examples are summarized in Table 1 for elastic resonance energy 17.674 MeV (22.129 MeV neutron energy).

**TABLE (1).** Maximum elastic time delay and speed in three examples

| Example, $\Gamma_d$, $\Gamma_n$ (MeV) | Time Delay, Speed (attosec) | Partial Width (0.5$\hbar$/col 2, MeV) |
|---|---|---|
| A) 0.0225, 0.0398 | 0.028, 0.020 | 0.012, 0.016 |
| B) 0.0083, 0.0303 | 0.071, 0.047 | 0.0047, 0.0070 |
| C) 0.0515, 0.0450 | 0.015, 0.008 | 0.022, 0.043 |

Example A) Deuteron and neutron partial widths 0.0225 and 0.0398 MeV are taken from multilevel *R*-matrix analysis and yield a scattering amplitude trajectory that passes above the center of the unitary circle. Corresponding phase shift increases through 90°.

Example B) Deuteron and neutron partial widths 0.0083 and 0.0303 MeV are taken from conventional pole time delay and speed plot distributions of example A. This example, distinguishing the two features of the resonance, is shown in Figure 1.

Example C) Deuteron and neutron partial widths 0.0515 and 0.0450 MeV are taken from a fit to Breit-Wigner model of the elastic resonant cross section of Example B convoluted with 0.040-MeV experimental energy spread. In this case, scattering amplitude trajectory passes beneath the center of the unitary circle and the corresponding phase shift does not increase through 90°.

Time delay and speed plots provide real parts of complex energies of *S*-matrix conventional (elastic) and shadow (reaction) poles 17.669±0.006 and 17.705±0.001 MeV n-α c.m. (0.046 and 0.082 MeV d-$^3$H c.m.), respectively.

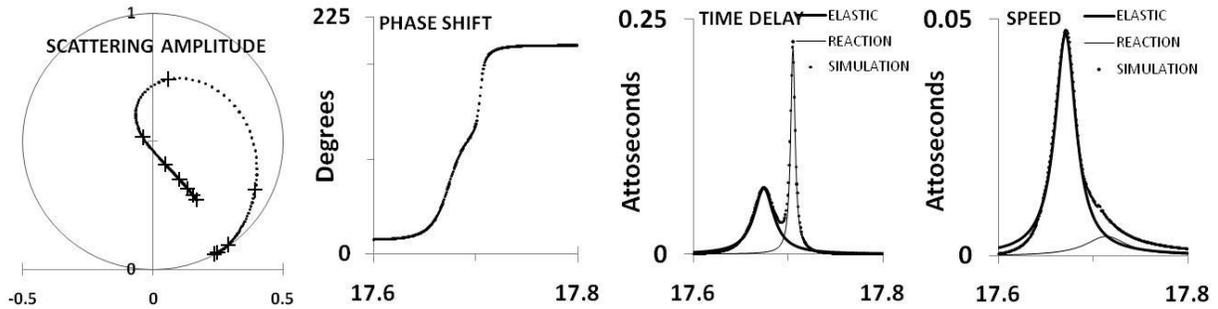

**FIGURE** 1. (Left two panels) Resonant $D_{3/2}$ scattering amplitude and corresponding resonant phase shift. Scattering amplitude trajectory, calculated at 0.001-MeV steps from 17.60 to 17.80 MeV n-α c.m. energy, encircles the center of the unitary circle. Large + symbols indicate values of n-α c.m. energy increasing (counterclockwise) in 0.020-MeV steps. Resonant phase shift plot vs n-α c.m. energy increases through 180°. (Right two panels) Corresponding real part of time delay and speed are compared with a best fit to the sum of two Lorentzians with (in the case of time delay) and without (in the case of speed) n-α relativistic transit time. Peak elastic resonance neutron energy is 22.129 MeV.

## SUMMARY AND CONCLUSIONS

In agreement with earlier observations [7], a Breit-Wigner approximation to the resonance suggests a relatively small deuteron partial width, compared to neutron width, which yields a narrow resonant elastic cross section. That is, Example B (Figure 1) suggests that the reported shadow pole width (ca. 0.008 MeV; cf. Ref.[2]) describes deuteron partial width, whereas the width (ca 0.030 MeV) of peak $^4$He(n,d)$^3$H reaction cross section relative to its maximum describes neutron partial width, as summarized earlier [7].

Accounting for 0.040-MeV experimental energy spread and cross section scale factor normalization, partial widths from maximum elastic time delay and speed plots of Example C taken as deuteron and neutron partial widths with resonance energy 22.124 MeV, yield behavior of Example A. As mentioned elsewhere [7], it must again be emphasized that all available elastic and total cross section data, uncorrected for 0.040-MeV experimental energy spread, are better represented by examples A and C, than by example B (Figure 1).

It must also be stressed that the *ab initio* method does not involve fitting experimental data. The present *ab initio* calculations employ similarity-renormalization-group-evolved NN potentials dependent on cutoff Λ. Therefore, characterization of resonant phase shift and energy may be different, by including three-nucleon (NNN) interaction or different NN potential with modified cutoff Λ.